\documentclass[letterpaper,12pt]{article}

\usepackage[T1]{fontenc}
\usepackage[utf8]{inputenc}
\usepackage{newtxtext,newtxmath} 

\usepackage[margin=1in,letterpaper]{geometry}
\usepackage{setspace}
\usepackage{microtype}   
\usepackage{parskip}     

\usepackage{amsmath}
\usepackage{gensymb}
\usepackage{tabularx}
\usepackage{graphicx}
\usepackage{float}

\usepackage{titlesec}
\titleformat{\section}{\Large\bfseries}{\thesection}{1em}{}
\titleformat{\subsection}{\large\bfseries}{\thesubsection}{1em}{}
\titleformat{\subsubsection}{\normalsize\bfseries}{\thesubsubsection}{1em}{}

\usepackage[numbers]{natbib}

\usepackage[final]{hyperref}
\hypersetup{
  colorlinks=true,
  linkcolor=blue,
  citecolor=blue,
  filecolor=magenta,
  urlcolor=blue
}

\begin{document}

\begin{titlepage}
\centering

\vspace*{1.5cm}

{\LARGE\bfseries
Analysis of the Ventriloquism Aftereffect\\
Using Network Theory Techniques
\par}

\vspace{1.5cm}

{\Large
Auditory Recalibration and Timeline of the\\
Ventriloquism Aftereffect
\par}

\vspace{2cm}

{\large\textbf{Sayan Saha}\par}

\vspace{1.5cm}

{\large
\textbf{Supervisor:}\\
Dr. Koel Das
\par}

\vspace{0.8cm}

Department of Mathematics and Statistics

\vspace{1.5cm}

\includegraphics[width=0.25\textwidth]{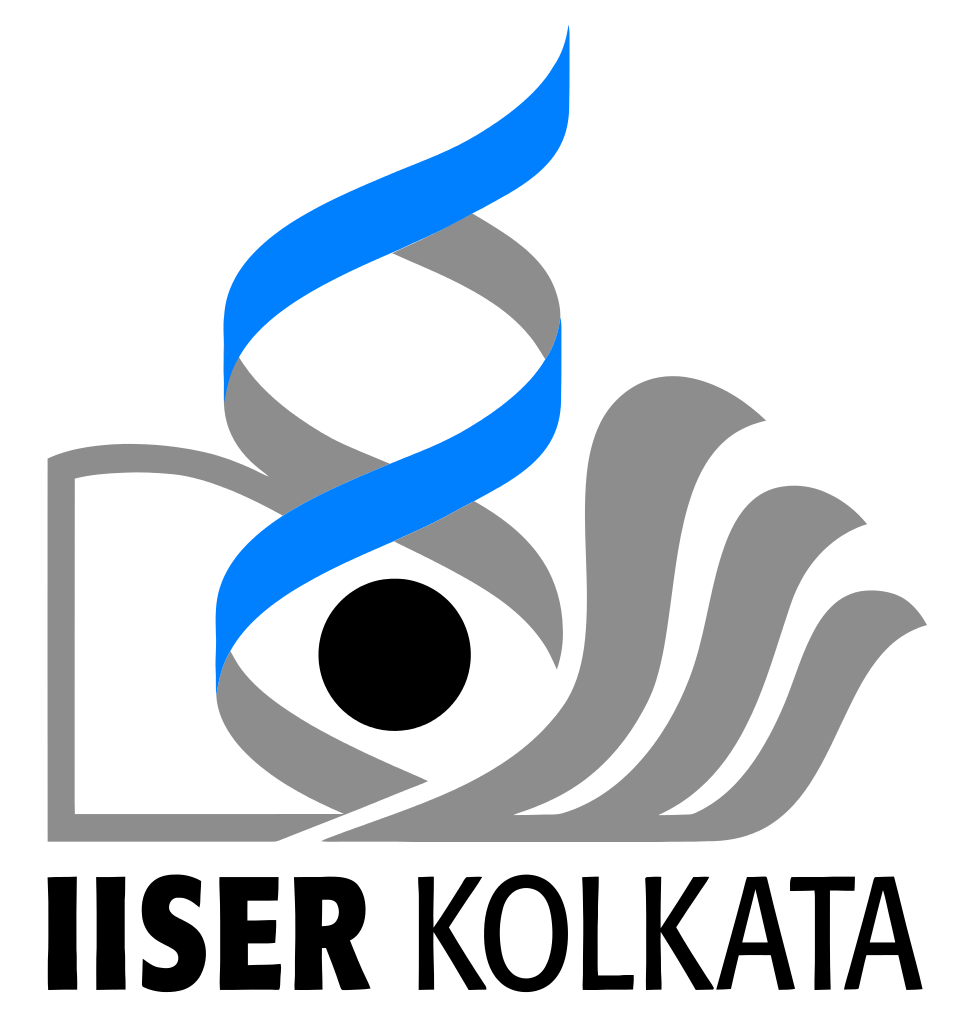}

\vfill

{\large
Submitted in partial fulfilment of the requirements\\
for the BS--MS degree
\par}

\vspace{0.5cm}

June 2020

\end{titlepage}


\section*{Declaration}
\addcontentsline{toc}{section}{Declaration}

I, Mr. \textbf{Sayan Saha} Registration No. 15MS032 dated August 1, 2015, a student of Department of
Mathematics and Statistics of the BS-MS Program of IISER Kolkata, hereby declare that
this thesis is my own work and, to the best of my knowledge, it neither contains materials previously
published or written by any other person, nor it has been submitted for any degree/diploma or any
other academic award anywhere before. I have used the originality checking service to prevent
inappropriate copying.

I also declare that all copyrighted material incorporated into this thesis is in compliance with the
Indian Copyright Act, 1957 (amended in 2012) and that I have received written permission from the
copyright owners for my use of their work.

I hereby grant permission to IISER Kolkata to store the thesis in a database which can be accessed by
others.

\vspace{1.5cm}

\textbf{Sayan Saha}\\
15MS032\\
Department of Mathematics and Statistics\\
Indian Institute of Science Education and Research\\
Kolkata\\
Mohanpur 741246, West Bengal, India

\newpage

\section*{Certificate from the Supervisor}
\addcontentsline{toc}{section}{Certificate from the Supervisor}

This is to certify that the thesis entitled "Auditory Recalibration and Timeline of the Ventriloquism After-Effect" submitted by Mr. Sayan Saha Registration No. 15MS032 dated August 1, 2015  a student of
Department of Mathematics and Statistics of the BS-MS Program of IISER Kolkata, is
based upon his/her own research work under my supervision. This is also to certify that neither the
thesis nor any part of it has been submitted for any degree/diploma or any other academic award
anywhere before. In my opinion, the thesis fulfils the requirement for the award of the BS-MS degree .

\vspace{1.5cm}

\textbf{Koel Das}\\
Associate Professor\\
Department of Mathematics and Statistics\\
Indian Institute of Science Education and Research\\
Kolkata\\
Mohanpur 741246, West Bengal, India

\newpage

\section*{Acknowledgement}
\addcontentsline{toc}{section}{Acknowledgement}

I thank my supervisor Dr. Koel Das who introduced me to the field of cognitive science research. She guided me and oversaw my efforts throughout the project while also giving me the independence to learn and apply other data analysis techniques throughout the scope of the project. I would also thank Dr. Satyaki Mazumder for helping me out with understanding the statistical tools involved for completion of the project and being the most approachable person and a mentor to me . I also thank my lab mates(Avishek Da, Tiasha Di and Adyasha) whom I reached out for help at various times during different stages of the project and making the lab a fun place to work. I must also thank Dr. Anirvan Chakraborty who guided me in the independent study courses parallel to my thesis where I was introduced to aspects of high dimensional statistics and reinforcement learning. I would always remember our conversations during my presentations in his office which were enlightening and eye-opening. I thank my dearest friends Saptarshi, Ron and Sweta for being the people they were during my life at the institute. I would warmly remember and cherish the memories we made. I also thank the institute and INSPIRE fellowship for supporting me during the duration of my BS-MS degree.

\newpage

\section*{Dedication}
\addcontentsline{toc}{section}{Dedication}

\begin{center}
\textit{ For my parents who have spent countless sleepless nights and held my hand, while I was navigating the turns of the serpentine paths of life.  I owe the completion of this long and arduous journey towards obtaining my Master's degree to them. }
\end{center}

\newpage

\begin{abstract}
Ventriloquism After-Effect is the phenomenon where sustained exposure to the ventriloquist illusion causes a change in unisensory auditory localization towards the location where the visual stimulus was present. We investigate the recalibration in EEG networks that causes this change and the track the timeline of changes in the auditory processing pathway. Our results obtained using network analysis, non-stationary time series analysis and multivariate pattern classification show that recalibration takes place early in the auditory processing pathway and the after-effect decays with time after exposure to the illusion.
\end{abstract}

\newpage
\tableofcontents
\newpage


\section{Introduction}
Ventriloquism is the illusion in which the voice appears to emanate from a puppet's mouth while it is actually the puppeteer.In laboratory settings, such an illusion can be reliably induced using synchronous visual and audio stimuli having a spatial disparity. Researchers have primarily looked at two effects of ventriloquism on subjects. In settings that involve bimodal trials it has been observed that there is a shift in the perceived location of the sound towards the source of the visual stimuli(ventriloquism effect). Also, in a training paradigm where a subject is exposed to a consistent spatial disparity between synchronously presented audio and visual stimuli, there is a recalibration of unisensory localization of sound in the direction of the visual stimuli following the training period. This kind of a recalibration is called the ventriloquism after-effect. We use dynamic network analysis techniques on EEG data taken during auditory localization tasks performed before and after ventriloqusim training to see how such recalibration is manifested in EEG time series networks.
\paragraph{}
Ventriloquism has often been studied as a tool to understand multisensory integration in the brain.  The effect of vision on auditory localization has been studied in various different experimental settings. In this section, we review a few notable studies that shed light on different aspects of the ventriloquism effect and ventriloquism after-effect. Temporal aspects of multisensory integration has come under the scanner in some studies especially when studying audio-visual integration. The speed of light is greater than the speed of sound by several orders of magnitude. However, the physiological auditory processing network is much faster than the network processing visual stimuli.Further, temporal effects are influenced by the degree of spatial overlap of the auditory and visual stimuli. A study \cite {slutsky2001temporal}where subjects were asked to distinguish between  distinct auditory and visual stimuli regardless of whether the stimuli occured synchronously found that responses were accurate regardless of the temporal disparity of the stimuli. However,with lesser spatial separation it became harder for the subjects to tell whether the stimuli occured at the same or different locations, with greater temporal disparity between the stimuli.
\paragraph{}
Soon after, studies investigating what makes the ventriloquist illusion weak or strong arrived at three crucial factors: the synchronicity of the stimuli , the spatial disparity between the stimuli and the power of the stimuli to captivate the subject. These factors interact in a complex manner to decide if the illusion overpowers our sensory modalities. Identification of these factors led to two different theories of sensory integration. The first, called the modality specificity hypothesis \cite{welch1980immediate,welch1999meaning},  proposes that the sensory modality with greater accuracy for the discrimination task at hand would dominate one's perception while performing the task. The visual system has an edge over the auditory system in this respect, hence the visual bias during the ventriloquism illusion can be explained successfully using this theory. Another theory uses the tool of Bayesian Probabilities to explain this integration\cite{burr2006combining,sato2007bayesian,alais2004ventriloquist}.
This theory supports the notion that the brain combines visual and auditory information in an effective way by minimizing variance. Since, the acuity of the visual stream is greater than the auditory stream, in the illusion, the brain gives greater weight to the visual stimuli during task performance.
\paragraph{}
The measurement of ventriloquism effect and the after-effect has to be done carefully. A distinction needs to be made between between immediate and cumulative recalibration effects. In an experiment paradigm using both bimodal and unimodal trials \cite{wozny2011recalibration}  reported at unimodal trials were influenced by the immediately preceding bimodal trial indicating  the immediate after-effect while the cumulative after-effect comes into effect after exposure to several bimodal trials.
The different timescales of this kind of a cross-modal recalibration has been the subjects of many study while some concluding that several hundreds of trials are needed for the recalibration to be manifested and a few reporting immediate recalibration effects. Theoretically, both mechaninsms can be explained by a single model\cite{bosen2018multiple}, a strong immediate aftereffect followed by a long tail that allows for accumulation of trials. Researchers have also reported different mechanisms for these different modes of recalibration\cite{bruns2015sensory,watson2019distinct}.
\paragraph{}
In, one of the landmark studies \cite{bertelson2000ventriloquist} investigating the role of deliberate, directed attention on the ventriloquism effect the authors concluded that the effect remains unchanged inspite of it. They noted that the effect is probably manifested due to its influence on our involuntary sensory modalities which dominate our voluntary cognitive modalities. However, later studies have reported top-down influences on the ventriloquism effect. In one study \cite{bruns2014reward} where high and low rewards were attached to bimodal stimuli, the task being localization of the sound source , the ventriloquism effect was significantly reduced for trials with high reward than trials with low reward, showing that a strong motivational goal of maximizing the reward can indeed take a toll on the effectiveness of the aftereffect.
A different line of study\cite{berger2013mental,berger2014fusion,berger2018mental} revealed that imagining a discrepant visual stimuli could induce identical ventriloquism after-effect further revealing top-down influences on cross-modal recalibration.
\paragraph{}
Many studies \cite{callan2015fmri,bonath2007neural,bonath2014audio}investigating the neural mechanisms of the ventriloquism effect have implicated the planum-temporale which is a space-sensitive region in the auditory cortex. Sensory accounts suggest that information from the auditory and visual system is integrated at an early stage, and the product of integration is available to our consciousness\cite{bakhman1999cognitive,kitagawa2002hearing,stekelenburg2004illusory,vroomen2001directing}.Cognitive accounts treat information from both the systems as independent modalities and suggest that the behavioural outcome is a result of decisional and cognitive biases\cite{alais2004no,meyer2001cross,sanabria2007perceptual,wuerger2003integration}. The ERP technique due to its high temporal resolution and f-MRI due to its high spatial resolution have been combined to effectively identify the biasing of left-right balance of the auditory cortex activity by spatially discrepant visual stimuli in the N260 component\cite{bonath2007neural}. Single neuron studies in animal models have implicated the superior colliculus for multisensory processing.
\paragraph{}
The ventriloquist illusion continues to be an effective tool for understanding multisensory integration.  Establishing  correspondence between computational models and neurobiological mechanisms would need designing of novel experiments aimed in this direction. The promise of a potential unifying theory of understanding multisensory processing in the brain awaits us in the near future.

\section{Methods}
\subsection{Network Analysis}
\subsubsection{Key Network Measures}
We review different network measures that characterize undirected networks with finitely many nodes in this section:
\paragraph{}
 \textbf{Degree of a Node} - It is the number of edges attached to the node.
\newline
Notation:- For all equations in the section the following notational convention shall be adopted:
\newline
$k_i$- Degree of node i
\newline
$p_k$- Probability that a randomly chosen node in the network would have degree k.
$L_i$- Number of links between the $k_i$ neighbors of node i
\newline
$A_{ij}$- No. of edges between node i and j 
\newline
$c_i$- Community the $i^{th}$ node belongs to
\paragraph{}
\textbf{Degree Distribution}- It is the probability distribution of
the degree of a given node in the network which is $p_k$ as a function of k.
\newline
\textbf{Local Clustering Coefficient}-  The local clustering coefficient of node i is the ratio of the number of connections between the neighbours of node i and the maximum number of edges between the neighbour of node i.
\begin{equation}
    CC_i=\frac{L_i}{k_i(k_i-1)}
\end{equation}
\textbf{Modularity}- Modularity is the fraction of edges that fall within a group minus the expected number of edges that would fall in the group if the edges were distributed in  random.
Let, 
\begin{equation}
    \sum_{i}k_i=2m
\end{equation}
Then, the modularity denoted by $Q$ is defined as
\begin{equation}
    Q=\frac{1}{2m}\sum_{vw}\left(A_{vw}-\frac{k_vk_w}{2m}\right)\delta(c_v,c_w)
\end{equation}
Participation Coefficient- It is the ratio between number of intermodular edges to the number of intramodular edges.
\subsubsection{Louvain Community Detection}
The problem of defining modularity of a network is that it depends on how the communities are defined. Hence, community detection becomes an important part in defining modularity. The Louvain Community Detection algorithm is a scheme that partitions the nodes into non-overlapping communities such that the modularity is maximized. The algorithm has two steps that are repeated iteratively:-
\newline
a) First each node is assigned to its own community. Then the change in modularity is calculated for each node \textbf{i} by removing it from its own community and moving it into the community of each neighbor of i one by one. Once this is done, \textbf{i} is placed in the community that resulted in the greatest modularity increase. If, no increase in modularity occurs then \textbf{i} remains in the community it previously was.
\newline
b) In the second step, all nodes from the same community is grouped together and a new network is built considering each community as a node. All  intracommunity edges are represented as self loops and and edges from multiple nodes in the same community to another node in a different community are represented as weighted edges between the modules
\paragraph{}
 It is worthwhile to note that  the value of modularity and the community structure obtained might not be the same in different runs of the algorithm. Although, one cannot reach a conclusion regarding the module structure, one can however sample the optimal modularity plateau a large number of times and use the average as the representative value for the network.

\subsection{Metrics}
\subsubsection{Phase Locking Value}
The performance of any cognitive function heavily relies on communication between different parts of the brain. There have been a lot of approaches in the literature that has tried to measure and effectively quantify  how such communication is established. These approaches were built on assumptions that empirical brain imaging experiments have validated, hence they have sound grounds. The measure Phase Locking Value had been proposed by Lachaux and his colleagues in 1999. The underlying assumption was that if two regions in the brain are working synchronously that is if they are functionally connected then the difference between the instantaneous phase between their signals must be constant. 
\paragraph{}
Hereby, we delve into the definition of instantaneous phase and how it can be extracted from a signal. The Fourier transform of a real valued signal is symmetric at 0 Hz, that is, both positive and negative frequencies are present. Although, negative frequencies make sense mathematically,  one cannot make sense of them physically. The most popular way to get around this problem in literature has been the use of analytic signals.
\begin{equation}
   z(t)=x(t)+iq(t) 
\end{equation}
where \textbf{x(t)} is the real valued signal
and \textbf{q(t)} is the Hilbert transform of \textbf{x(t)}.
\begin{equation}
    q(t)=\frac{1}{\pi}P\int_{-\infty}^{+\infty}\frac{x(t)}{t-t^\prime}dt^\prime
\end{equation}
where P indicates the Cauchy Principal Value.
The instantaneous phase is defined as the angle between the real and imaginary part of the analytic signal \textbf{z(t)}. For this defintion to make sense in a physical context the signal has to be narrowband filtered. Then, the phase locking value between two different signals can be computed as:
\begin{equation}
    PLV(t)=\frac{1}{N}|\sum_{n=1}^{N}e^{i\theta(t,n)}|
\end{equation}
where, N is the number of trials and $\theta(t,n)$ is difference between the instantaneous phase of the two signals at timepoint t and trial n. It is clear from the definition that PLV only takes on non-negative values. It is also worthwhile to note that the range of values is in the interval [0,1]. A PLV of 1 indicates absolute phase synchrony whereas a PLV of 0 indicates no phase synchrony.
\newline
\textbf{Volume Conduction:-}
\newline
One of the major problems of using connectivity measures on scalp space data is the problem of volume conduction. The nodes of the network we are trying to build are electrodes. We want the edges between these nodes to be a good representative of interactions between brain regions lying directly underneath the electrodes. However, this might not  always be the case. A common source inside the brain might influence two separate electrodes giving rise to spuriously high correlations or phase synchrony between these electrodes. The next connectivity measure tries to overcome this problem. 
\subsubsection{Imaginary Part of Coherence}
Coherence has been the most common measure used to investigate connectivity between signals(EEG/MEG) from two different brain regions. It tries to quantify the linear relationship between two signals at a given frequency. It can be calculated as:
\begin{equation}
    Coh(f,t)=\frac{\sum_{n=1}^{n=N}S_1^n(f,t)S_2^{n*}(f,t)}{\sqrt{|\sum_{n=1}^{n=N}S_1^n(f,t)|^2|\sum_{n=1}^{n=N}S_2^n(f,t)|^2}}
\end{equation}
where N is the total number of trials, $S_1^n(f,t)$ and $S_2 ^n(f,t)$ are short term fourier transformed signals from electrodes 1 and 2 respectively at frequency f and timepoint t.
\newline
Another way to define coherence is through the cross-spectrum. Let $x_i(f)$ and $x_j(f)$ be the Fourier transforms of time series $x_i()t$ and
$x_j(t)$. Then the cross-spectrum is defined as
:
\begin{equation}
    S_{ij}(f)=<x_i(f)x_j^*(f)>
\end{equation}
where, <> denotes the expectation operator. Coherency can now be defined as:
\begin{equation}
    C_{ij}(f)=\frac{S_{ij}(f)}{\sqrt{S_{ii}(f)S_{jj}(f)}}
\end{equation}
Coherence just like PLV, tries to quantify the phase synchrony between two signals. Hence, it is worthwhile to compare certain facets of it with the Phase Locking Value.  Lachaux et. al has argued in his paper that Phase Locking Value is a better way to interpret phase locking. The argument in favor of PLV centers on the fact that phase locking is independent of amplitudes hence is a better measure of phase relationship.
However, it is not clear whether statistical independence of phase relationships from amplitudes is actually the case. Another argument in favor of PLV centers on the fact that coherency can only be applied to stationary signals.  This is true when coherency is interpreted as a parameter of a  stationary process which we would not do here.  Similarly quantifying linear relationships does not mean that we assume there are no non-linear relationships. 
\newline
Although, it is a valid and a very popular measure of phase synchrony, coherency by itself cannot overcome the problem of volume conduction. Nolte et.al.\cite{nolte2004identifying} in 2004 proposed  that the imaginary part of coherency can overcome the problem of volume conduction. A simple approach to quantify effects of volume conduction  on coherency show that effects of common sources can only manifest itself on the real part of coherence, the imaginary part is independent of such effects. Let us assume signals in electrodes i and j are caused by linear superposition of K common sources in  the brain ,call them $s_k$. 
\begin{equation}
    x_i(f)=\sum_{k=1}^{K}z_{ik}s_k(f)
\end{equation}
\begin{equation}
    x_j(f)=\sum_{k=1}^{K}z_{jk}s_k(f)
\end{equation}
Now, we compute the cross spectrum:
\begin{equation}
\begin{split}
    S_{ij}(f)= <x_i(f)x_j^*(f)>\\
    =\sum_{kk\prime}z_{ik}z_{jk\prime}<s_k(f)s_{k\prime}^*(f)>\\
    =\sum_{k}z_{ik}z_{jk}<s_k(f)s_{k}^*(f)>\\
   =\sum_{k}z_{ik}z_{jk}|s_k(f)|^2
\end{split}
\end{equation}
which is real.  Since, the denominator in coherency is also real, it follows hence that coherency is real. The proof rests on some basic assumptions. It assumes a linear superposition of sources which is justified due to the linearity of the Maxwell's equation. Another subtle assumption is that the mapping between sources and electrodes is free of phase shifts. This assumption is justified because a phase shift in the frequency domain is a time lag in the time domain and a signal cannot be time-lagged to itself. 
\subsection{Empirical Mode Decomposition}
Data from neuroimaging experiments just like most other physical experiments poses challenges to data analysers attempting to make sense out of it. A few of them are: a) the data span may be too short b) the data is non-stationary c) the data arises from non-linear processes. Tools to analyse such data in literature are difficult to come by. Fourier spectral analysis is the traditional tool for obtaining a energy distribution spectrum from time series data. Methods such as short term fourier transform are  very commonly used to obtain a time-frequency spectra from time series data. We discuss the limitations these methods suffer from. The Fourier transform works in very general scenarios but it has some crucial limitations.  If the system is not linear and the data is not stationary, it makes little  physical sense. Firstly, the Fourier spectrum defines globally uniform harmonic components, needing additional components to simulate non-stationary data. This effect is manifested as the spread of energy over a broad frequency range in the spectrum.
We briefly discuss the exact drawbacks of a couple of popular methods used for data analysis in EEG experiments that rely on Fourier transform.
\newline
a)\textbf{Spectrogram}:-
\newline
Spectrogram is exactly the same as Fourier Transform with a limited time window width that is segmenting the time series into non-overlapping parts and taking the Fourier transform of each of them. One takes a sliding window approach to obtain a time-frequency distribution. One has to assume piecewise stationarity to adopt this approach. Even if the data is piecewise stationary, there is no guarantee that the window size adopted coincides with the stationary time scales. Even if we somehow manage to coincide the window width with the stationary time scale, it is not clear what can we know about the data beyond the stationary time scale.
\newline
b)\textbf{Wavelet Analysis}:-
\newline
Wavelet is essentially an adujustable window 
Fourier spectral analysis with the definition:-
\begin{equation}
    W(a,b;X,\psi)=|a|^{-\frac{1}{2}}\int_{-\infty}^{\infty}X(t)\psi^*\left(\frac{t-b}{a}\right)dt 
\end{equation}
where $\psi^*()$ is the basic wavelet function, a is the scale factor and b denotes the translation from the origin. Time and frequency are implicitly embedded in the definition, the variable $\frac{1}{a}$ gives the frequency scale and b denotes the temporal location of the event. $W(a,b;X,\psi)$ defines the energy of X at scale a and timepoint b. In general $\psi^*$ is not orthogonal for different a for continuous wavelet functions. Although, choosing a discrete set of a makes the wavelets orthogonal, this discrete wavelet analysis misses signals with scales different from a.  Wavelet analysis provides a uniform resolution for all scales however the downside of uniform resolution is a uniformly poor resolution. Interpretability of wavelet is often counter-intuitive. For example, to define a local change one must look at higher frequency ranges for higher the frequency more  localized the basic wavelet would be. Hence,  to detect a local event occurring in a low frequency range one would be forced to look at higher frequency ranges.
\paragraph{}
The above methods designed to modify Fourier spectral analysis for a global representation fail in one way or another. The necessary conditions for a basis to represent a non-stationary and non-linear time series are :- a)complete b)orthogonal c)local d)adaptive.  The first and second condition guarantee precise decomposability without leakage. Requirement of locality is important for non-stationarity as such data does not have an intrinsic time scale. Adaptivity takes into account the local variations of the data. Hence, amplitude and frequency are both functions of time in such a basis. In other words we need the instantaneous frequency and energy rather than global frequency and energy as defined by the Fourier based methods.
\paragraph{}
\textbf{Instantaneous Frequency:-}
\newline
Traditionally in Fourier analysis frequency is defined using a sine or cosine function spanning the whole data length with constant amplitude. To be coherent with this definition, the definition of instantaneous frequency needs to be defined using a sine or cosine function  which needs a full wave span but such a definition would not make sense for  non-stationary data where frequency changes with time.  The second difficulty in defining instantaneous frequency arises from the non-uniqueness of the definition. However, that has been resolved with the introduction of Hilbert Transform. If x(t) is an arbitrary time series, then,
\begin{equation}
    q(t)=\frac{1}{\pi}P\int_{-\infty}^{+\infty}\frac{x(t)}{t-t^\prime}dt^\prime
\end{equation}
where P indicates the Cauchy Principal Value. q(t) is the hilbert transform of x(t). Using the hilbert transform, we define the analytic signal z(t)  as:
\begin{equation}
    z(t)= x(t) + iq(t) = a(t)e^{i\theta(t)}
\end{equation}
where
\begin{equation}
    a(t)= (x(t)^2 + q(t)^2)^{\frac{1}{2}}
\end{equation}
and
\begin{equation}
\theta(t)=arctan\left(\frac{q(t)}{x(t)}\right)
    \end{equation}
The polar coordinate expression where $\theta(t)$ is the instantaneous phase clarifies the local nature of the definition; it is the best fit of an amplitude and phase varying trigonometric function to x(t). Thus, ideally instantaneous frequency should be defined as:-
\begin{equation}
    \omega(t)=\frac{d\theta(t)}{dt}
\end{equation}
However, for $\omega(t)$ to be a single valued function of time, one needs to impose some more restrictions on the time series x(t). Such a signal for which instantaneous frequency can be clearly defined has been called a 'monocomponent' signal in the literature. However,  a rigorous definition of a monocomponent signal is never clearly mentioned although various kinds of global restrictions have been imposed on signals in literature to obtain a well defined notion of instantaneous frequency. Two necessary local restrictions on a time for which a well defined instantaneous frequency can be obtained are
:a) it has to be symmetric with respect to the local zero mean b)it must have the same number of zero crossings and extrema.  These restrictions inspire the definition of intrinsic mode function. An intrinsic mode function(IMF) is a function having the properties:- a) in the whole data set the number of zero crossings and number of extrema must be either equal or differ by almost one b) locally the mean value of the envelope defined by the local maxima and the envelope defined by the local minima is zero. It should be noted that an IMF can be amplitude and frequency modulated, in fact it can be non-stationary. The Hilbert transform of an IMF is well-behaved in the sense that it allows to define instantaneous frequency uniquely as a single valued function of the phase.
\newline
\textbf{The Sifting Process:-}
\newline
Most of the data one encounters in physical measurements
are not IMFs . Hence, one has to decompose the data into IMFs. The decomposition is based on the following assumpitons: a) the data has atleast two extrema- a minimum and a maximum. Even, if it is completely devoid of extrema it can be differentiated one or more times to reveal the extrema. b) the intrinsic time scale is defined by the time scale of the extrema. 
\newline
The decomposition method uses the local maxima and minima separately. Once the extremas are identified, the local maximas are connected using a cubic spline to form the upper envelope and the same procedure is repeated for the local minimas to form the lower envelopes. The mean is designated as $m_1$ and the difference between the data and $m_1$ is the first component $h_1$ ,i.e,
\begin{equation}
    X(t)-m_1=h_1
\end{equation}
Ideally, $h_1$ should satisfy the requirements of an IMF, but in realistic cases it often does not. Overshoots and undershoots generate new extrema and exaggerate  existing ones. Another complication due to non-linearity might be that the envelope mean might be different from the true local mean. The sifting process aims to eliminate riding waves and to remove asymmetric components from wave profiles. In the second iteration of the sifting process $h_1$ is treated as the data. If carried out without a meaningful stopping criterion, the sifting process can eliminate meaningful amplitude fluctuations. A stopping criterion can be devised by setting a tolerance criterion on the standard deviation between two successive sifting results:  
\begin{equation}
    SD=\sum_{t=1}^{T}\frac{(h_{1k-1}(t)-h_{1k}(t))^2}{h_{1k-1}^2(t)}
\end{equation}
A value of SD can be set between 0.2 and 0.3. The first IMF
component $c_1$ obtained after stopping the sifting process according to the tolerance criterion should contain the finest scale or the fastest oscillatory mode of the signal. One can now separate this component from the signal:-
\begin{equation}
    X(t)- c_1= r_1
\end{equation}
$r1$ contains information about longer period components of the signal. It can be treated as new data and subjected to the same sifting process to get the next component $c_2$.
The procedure can be repeated on all the subsequent $r_i's$
and the result is:-
\begin{equation}
    r_1- c_2= r_2....r_{n-1}-c_n=r_n
\end{equation}
This process can be stopped when the residue $r_n$ is small enough to be inconsequential or a monotonic function from which no more IMF can be extracted. This process achieves a decomposition of the data into n empirical modes and a residual.
\begin{equation}
    X(t)=\sum_{i=1}^nc_i+r_n
\end{equation}. After performing the Hilbert transform, on each IMF component the data can be expressed in the following form:-
\begin{equation}
    X(t)=\sum_{k=1}^{n}a_k(t)e^{i\int\omega_k(t)}
\end{equation}.
The same data if represented in Fourier expansion would be:
\begin{equation}
    X(t)=\sum_{k=1}^{n}a_ke^{i\omega_kt}
\end{equation}.
Equation (24) gives amplitude and frequency as a time varying function whereas in Equation(25) they are constants. The IMF expansion accomodates the non-stationarity of the data allowing time variability of amplitude and frequency hence improving the time-frequency resolution.
\section{Experiment and Results}
\subsubsection{Experimental Paradigm}
The experiment was conducted by Dr. Brigitte Roder and her colleagues\cite{bruns2011cross}. We use EEG data from this experiment for our analysis. Subjects performed an auditory localization task before and after a ventriloquism training period. The task consisted of four sessions of 114 trials each.  In each trial an auditory beep of 2000 Hz was emitted from one of the three speakers located at -10\degree, 0\degree and +10\degree relative to the subject's head (negative values indicate location to the left of the observer's head). After, a random delay(700 ms- 900ms), a go signal was presented from all the three speakers, after which the subjects had to report the perceived location(left,center or right) of the auditory beep. The delay was introduced to prevent contamination of EEG due to motor related artifacts. The beeps in each session was presented from each speaker equal number of times in a random manner. After, the pretest period the participants were subjected to sustained ventriloquism training.
\paragraph{}
The training stimuli comprised synchronously presented light(yellow LED) and auditory stimuli(identical to the task stimuli) with a constant spatial discrepancy. Auditory stimuli were presented at the same three locations as the task with the light stimuli offset by 15\degree to the left or to the right of the auditory stimuli depending on the participants group. Half of the participants were trained on light stimuli to the left of the auditory stimuli (left-adapted) and half were adapted to the light stimuli to the right of the auditory stimuli (right-adapted). The training stimuli was presented in sets of five consecutive stimuli from the same location at a rate of 1 per second. Equal number of stimuli was presented from each location. Participants had to respond to occasional deviant stimuli(either a red LED light or a 1000 Hz tone) to ensure that they were attentive during the training.
\subsubsection{EEG Modalilties}
EEG was recorded using 60 Ag/AgCl electrodes organized according to the 10-10 system. All channels were referenced to the average of left and right earlobe recordings. Blinks and vertical eye movements were measured with an electrode beneath the right eye. Horizontal eye movements were detected using electrodes placed in the left and right outer canthi. The data was sampled at 500 Hz and filtered offline with a high cut off of 40Hz using a Butterworth filter of order 5. 
\subsubsection{Spectral Analysis}
Spectral decomposition of EEG signals was carried out using the short term Fourier transform method. (spectrogram function in MATLAB). The time window length adopted was of 20 timepoints and the window overlap was 25\%. This estimation was done for 40 linearly spaced frequency points. Hence for every subject, condition and channel we obtained a 3D matrix of number of frequency points x no. of time points x no. of trials. 
\subsubsection{Functional Connectivity Estimation}
We have used imaginary part of coherence(Imcoh) and Phase Locking Value(PLV) as the measure of functional coupling. Imaginary part of coherence is a measure insensitive to volume conduction. Using each coupling measure, for every subject, condition, time window and frequency point we obtained a adjacency matrix of the order of the no. of electrode pairs(60x60). For each subject we had 6 conditions depending on  the when the task was being performed(pre-training or post training)  and the location of the auditory stimulus(left, right or center).
\subsubsection{Network Analysis}
We firstly studied differences in nodal clustering coefficients between the pre-training and post-training networks. 
\begin{figure*}[htp]
\caption{Electrodes where nodal clustering coefficient is significanty greater ($p<0.05)$ in right adpated audio left conditions in the beta band using Imcoh}
\centering
\fbox{\includegraphics[scale=0.85]{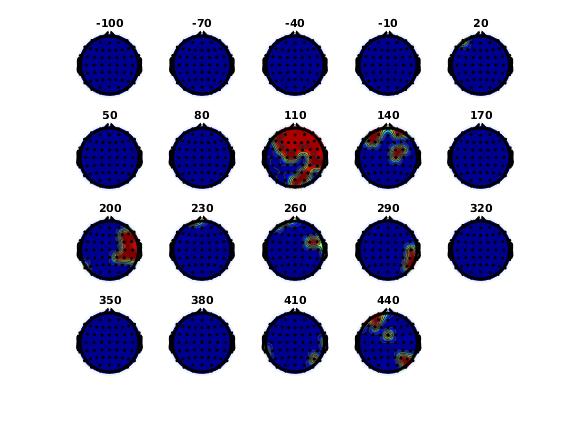}}
\end{figure*}
For networks obtained using imaginary part of coherence, We found that the nodal clustering coefficient is significantly greater in the post-training than the pre-training networks($p<0.05$)  especially in the 100ms-150ms time window for right adapted subjects during localization tasks where the sound was emitted from the speaker to their left(audio-left condition) and for left adapted subjects when  the sound was emitted from the speaker to their right in the alpha, beta and the theta band(audio-right condition). However, after Bonferroni Correction and correction using False Discovery Rate the difference was not significant. The difference in clustering coefficient between the pre-training and post-training for the audio-center condition in both the right-adapted and left-adapted subjects was confined to time windows  before the stimulus presentation in the alpha and beta band and time windows late after stimulus in the theta band. The differences were not significant after Bonferroni correction. 
\paragraph{}
For networks obtained using Phase Locking Value, we found that the nodal clustering coefficient is not significantly different in the post-training than the pre-training networks  in  any time window for right adapted or left adapted subjects during localization tasks in any condition( audio left , audio center or audio right) in any frequency effect.

\paragraph{}
We also studied the community structure in the pre-training and post-training networks. We investigated the modularity of these networks to gain insight about the community organization. Importantly, investigating differences in community structure between conditions would be relevant if the networks we obtained were indeed modular when compared to null models. We created 25  null network models(Newman- Girvan null models) for any given network by destroying any possible modular structure in the original network by randomizing edges while preserving the degree, weight and strength distributions. Each null network was partitioned into communities using the same procedure as the original networks. Difference between modularity of the original network and the null networks was computed and tested against 0.  All of the networks we obtained were significantly modular($p<0.05$), in line with previous results obtained for cognitive networks.
\paragraph{}
In order to define modularity uniquely for the networks, we used a greedy optimization method for community detection. The Louvain Community Detection Algorithm was used to optimize modularity($Q_{uni}$) and partition the network into non-overlapping communities. Following several iterations of this algorithm, the high modularity plateau was extensively sampled and measures of interest were computed for each community assignment in this plateau and averaged over all community assignments to get a representative measure for the underlying network.

\begin{figure*}[htp]
\caption{No. of modules  vs time graph averaged across left-adpated subjects in the audio right  condition in alpha band networks using Imcoh  }
\centering
\fbox{\includegraphics[scale=0.5]{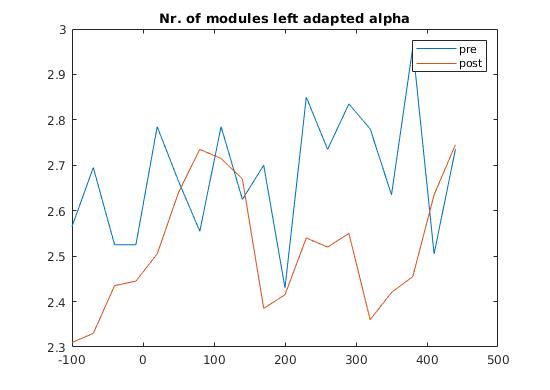}}
\end{figure*}
\newpage
\begin{figure*}[htp]
\caption{ Modularity vs time graph averaged across left-adpated subjects in the audio right  condition in alpha band networks using Imcoh}
\centering
\fbox{\includegraphics[scale=0.5]{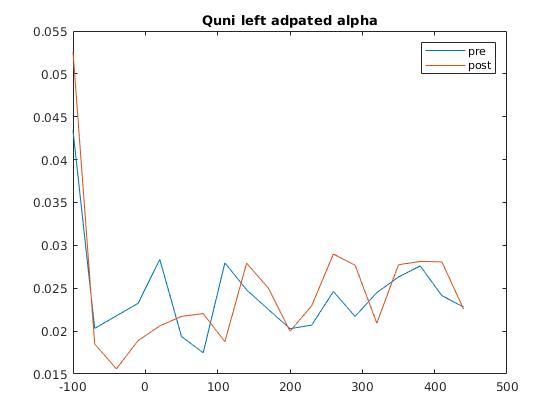}}
\end{figure*}
\paragraph{}
The simple features we analysed for community orgainization analysis were: a) Modularity b)Number of modules. These features were computed separately for each frequency band and each time-point and averaged across subjects. We did not find significant difference in both these features owing to these measures being coarse in their very nature and great variance across subjects. This prompted us to use finer measures for analysis of modular structure in order to identify nodes that might possibly behave as hubs in these networks and nodes that might play a crucial role in intramodular connectivity.
\newline
It is worth noting here that Phase Locking Value not being immune to volume conduction effects implies that spatially proximal electrodes would have links between them with higher weights which would result in a community structure where electrodes adjacent to each other would be in the same module, resulting in a very low number of modules( typically two encompassing each half of the scalp).
This would fail to reflect  connectivity of underlying sources in the brain. Hence, it would not be meaningful to investigate  community structure in networks obtained using Phase Locking Value.
\paragraph{}
For networks obtained using imaginary part of coherence, the participation coefficients of centro-parietal electrodes was significantly greater during the post-training task around the 100 ms time window in the alpha-band for the audio-left and audio-right condition in the right -adapted subjects. However, the effects were not significant after Bonferroni correction.  For left adpated subjects the participation coefficients of the  centro-parietal was significantly greater during the post-training task around the 100 ms time window in the alpha-band for the audio-right condition.However, the effects were not significant after Bonferroni correction.
This shows that atleast a part of recalibration necessary for the ventriloquism after-effect takes place due to the centro-parietal regions acting as hubs supporting intermodular connectivity. 

\section{Classification of EEG Data}
We have also tried using automated classification algorithms to differentiate pre-training and post-training data. We have used a classification algorithm called the CPCA( classwise Principal Component Analysis)\cite{das2009efficient}. This pattern recognition algorithm has the advantage of discarding non-informative subspaces in the data, giving it an edge when working with small sample number and high dimensional data.
\paragraph{}
We classified the data subject wise and used several kinds of approaches to find different kinds of distinguishing features of the data. We tried to investigate the differences in EEG data between a sound and its ventriloquized counterpart while the location from which the auditory stimuli emitted was fixed for each subject. First, each input data to the classifier was the entire trial time series data for all the electrodes. The dimension of data( no. of electrodes x no. of timepoints) for this classifier was very high. The number of samples for each class corresponded to the number of trials the subject underwent for that auditory stimuli depending on the class. Ten fold cross validation was performed. The classification performance was measured by looking at the PC and Wilcoxon signed rank test was performed to check if the classification performance was higher than chance. The mean PC for right adpated subjects for the audio-left condition was 0.5658 significantly higher than chance ($p<0.01$).The mean PC for right adpated subjects for the audio-center condition was 0.5501 significantly higher than chance ($p<0.01$).The mean PC for right adpated subjects for the audio-right condition was 0.5650 significantly higher than chance ($p<0.01$).
The mean PC for the left adapted subjects for the audio-left condition was 0.5251 significantly higher than chance($p<0.04$). The mean PC for the left adapted subjects for the audio-center condition was 0.5247 significantly higher than chance($p<0.04$).The mean PC for the left adapted subjects for the audio-right condition was 0.5279 significantly greater than chance($p<0.02$).
An important aspect of the classification rate is that in both the left and right adapted subjects the classification rate in the audio-center condition is lower than the other conditions. Also, the classification rate for the left-adapted subjects is highest in the audio-right condition and the classification rate for the right adapted subjects is highest in the audio-left condition. 

\paragraph{}

The same classification was performed with the aid of Empirical Mode Decomposition. We used raw EEG data with the baseline removed for Empirical Mode Decomposition. The input to the classifier was the first Intrinsic Mode Function obtained from decomposition of the time series data for each electrode. The number of features of the classifier was identical to the last setting. Ten fold cross validation was performed. The classification performance was measured by looking at the PC and Wilcoxon signed rank test was performed to check if the classification performance was higher than chance. The mean PC for right adpated subjects for the audio-left condition was 0.5643 significantly higher than chance ($p<0.01$).The mean PC for right adpated subjects for the audio-center condition was 0.5501 significantly higher than chance ($p<0.01$).The mean PC for right adpated subjects for the audio-right condition was 0.5630 significantly higher than chance ($p<0.01$).
The mean PC for the left adapted subjects for the audio-left condition was 0.5351 significantly higher than chance($p<0.04$). The mean PC for the left adapted subjects for the audio-center condition was 0.5238 significantly higher than chance($p<0.04$).The mean PC for the left adapted subjects for the audio-right condition was 0.5249 significantly greater than chance($p<0.02$).

\paragraph{}
We also hypothesised that the ventriloquism after-effect would decay with the number of trials the subject underwent after training.  In order to validate our hypothesis, we performed a classification using "early trial" data. First 50\% trials after each training session for a subject were labelled as early trials. In a setting, identical to the previous classifier, the time series from these early trials were used as input data.The number of features of the classifier was identical to the last setting. Ten fold cross validation was performed. The classification performance was measured by looking at the PC and Wilcoxon signed rank test was performed to check if the classification performance was higher than chance. The mean PC for right adpated subjects for the audio-left condition was 0.5541 significantly higher than chance ($p<0.05$).The mean PC for right adpated subjects for the audio-center condition was 0.5443 significantly higher than chance ($p<0.05$).The mean PC for right adpated subjects for the audio-right condition was 0.5556 significantly higher than chance ($p<0.05$).
The mean PC for the left adapted subjects for the audio-left condition was 0.5459 significantly higher than chance($p<0.05$). The mean PC for the left adapted subjects for the audio-center condition was 0.5233 significantly higher than chance($p<0.05$).The mean PC for the left adapted subjects for the audio-right condition was 0.5251 significantly greater than chance($p<0.05$).
The classification performance being identical to the classifier where all trials were used indicates that most of the discriminatory information is present in the early trials as the after-effect decays with time lapse after training.
\begin{figure*}[htp]
\caption{Classification Performance in Right Adapted Subjects }
\centering
\fbox{\includegraphics[scale=0.5]{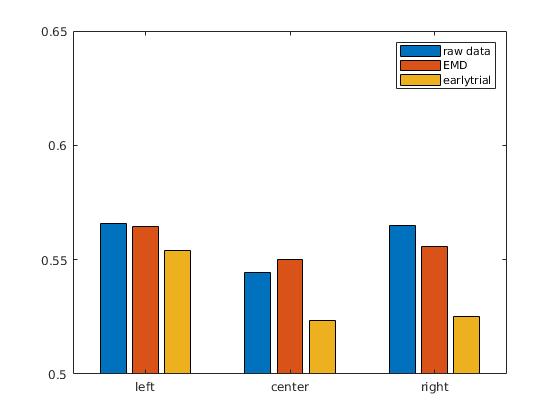}}
\end{figure*}
\newpage
\begin{figure*}[htp]
\caption{Classification Performance in Left Adapted Subjects }
\centering
\fbox{\includegraphics[scale=0.5]{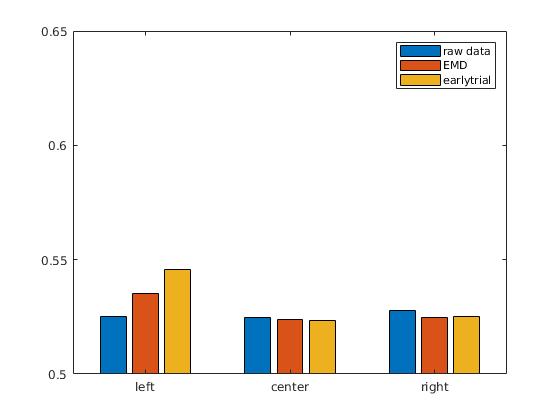}}
\end{figure*}
Next we tried,classifying the data time-window wise, again keeping the location of the auditory stimuli fixed  to check for temporal trends in classification performance. The idea behind this was to check in which time window the classes are best distinguishable, which would correspond roughly to the time in the auditory processing pathway when a sound and its ventriloquized counterpart appears different to the brain. The input data to the classifier was the time-averaged data for the particular time window,  for each electrode. Hence, the dimension of the data was equal to the no. of electrodes, which was considerably less. We performed ten fold cross validation in each time window. We used the Wilcoxon signed rank test to determine if the PC was significantly higher than chance.
\begin{figure*}[htp]
\caption{Classification Performance trend in Left Adapted Subjects }
\centering
\fbox{\includegraphics[scale=0.45]{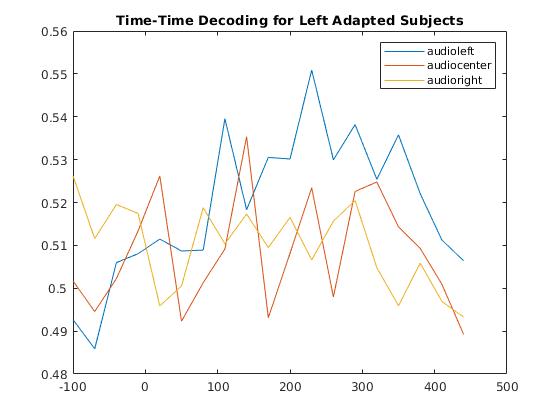}}
\end{figure*}
The classification performance in both left and right  adapted subjects were significantly above chance in each of the time windows between 100ms to 260ms for the audio-left and audio-right conditions, however the classification rate was not significantly higher than chance in these time windows for the audio-center condition. The peaking  of classification performance during the 100ms-260ms time range depicts that a sound and its ventriloquized counterpart is differentiated by the auditory processing pathway in this time range. 
\begin{figure*}[htp]
\caption{Classification Performance trend in Right Adapted Subjects }
\centering
\fbox{\includegraphics[scale=0.45]{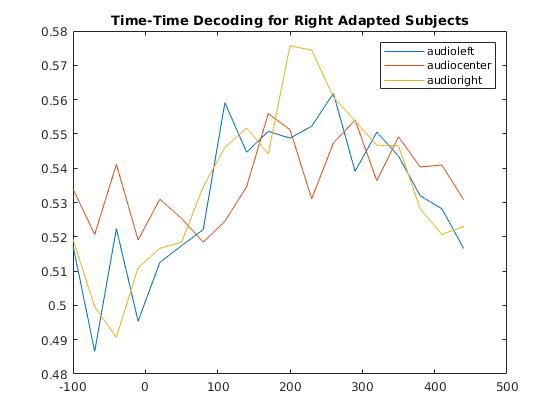}}
\end{figure*}

\newpage
\section{Conclusion}
Ventriloquism is indeed a phenomenon that can be studied using independent approaches to shed light on different neural mechanisms during audio-visual cognition. Our study reveals two aspects of ventriloquism after-effect through  network analysis and classification tools:-  a)the brain distinguishes between a sound and its ventriloquized counterpart at an early stage of the auditory processing pathway b) the after-effect decays with trials after training.  A natural extension of this study would be to use networks obtained after Hilbert-Huang transform of EEG data since it gives a finer temporal resolution or use f-MRI data and correlation measures  to implicate the sources responsible for such recalibration.
\paragraph{}
Another direction one can pursue while investigating ventriloquism effect is the notion of the brain as a predicitive coding machine\cite{clark2013whatever}. This recent development views the brain as an organ that guides one's perception of the environment based on predictions of the internal models it forms of the surroundings. According to this theory, the brain updates its internal models when its internal model makes error in the prediction process. Hence in this setting, cognition involves a bottom up transmission of sensory impulses and a top down transmission of prediction errors. The recalibration during ventriloquism can be viewed as a recalibration in the internal model of the environment and the visual prior dominating the auditory impulse. 
\paragraph{}
Ventriloquism being an illusion offers a lens to investigate implicit neural biases that our information processing systems in the brain have. Questions  in the domain of how our brain processes conflicting sensory stimuli and how priors in our internal model of the environment change in the presence of such conflicting stimuli, become more important than ever as researchers try to understand mechanisms that bias our behaviour at a neural level leaving little room for us to fight for cognitive conscious self to overcome such bias.


\newpage
\bibliographystyle{unsrtnat}
\bibliography{ref} 
\end{document}